\documentclass[aps,twocolumn,
showkeys,showpacs,nofootinbib,byrevtex]{revtex4-1}
\usepackage{amsmath,amsfonts,amssymb,amscd,amsxtra,amsthm}
\usepackage{graphicx}
\usepackage{bm}
\usepackage{ulem}
\usepackage{epsfig}

\begin{document}

\title{Role of $\sigma$ exchange in the $\gamma p\to \phi  p$ process
and scaling with the $f_1$ axial vector meson from a Reggeized
model}

\author{Byung-Geel Yu}%

\email{bgyu@kau.ac.kr}%

\author{Hungchong Kim}%


\author{Kook-Jin Kong}%
\email{kong@kau.ac.kr}%
\affiliation{Research Institute of Basic Science, Korea Aerospace
University, Goyang, 412-791, Korea}


\begin{abstract}
We investigate the role driven by the scalar meson $\sigma$
exchange in the photoproduction of the vector meson $\phi$(1020)
off a proton by using a Reggeized model. Based on the
$\pi^0(135)+\sigma(500)+f_2(1270)$+Pomeron exchanges, we
demonstrate that the $\sigma$ exchange plays the role to reproduce
the bump structure at the forward angle in the differential cross
section as well as the peaking behavior in the total cross section
observed in the CLAS Collaboration.  We also discuss the possible
observation of the scaled cross section $s^7d\sigma /dt$ at the
production angle $\theta=90^{\circ}$ from the CLAS data. It is
found that the axial vector meson $f_1(1285)$ exchange with the
trajectory $\alpha_{f_1}(t)=0.028\,t+0.9\pm0.2$ arising from the
axial anomaly of the QCD vacuum plays the role to clarify the
scaling up to 5 GeV.
\end{abstract}

\pacs{11.55.Jy, 13.40.-f, 13.60.Le}

\keywords{$\phi$ vector meson, photoproduction, scaling, scalar
meson, axial vector meson}

\maketitle

Photoproduction of the neutral vector meson has been an important
tool to explore QCD dynamics via the hadronic process. Especially,
the diffractive feature of the reaction process showing at high
energies has drawn attention  for decades, and such a nonmesonic
$t$-channel peripheral process is, to date, materialized by the
Pomeron exchange \cite{laget}.

Among the light vector mesons, the $\phi$ meson photoproduction
off the proton target is special because the physical  $\phi$
meson is a pure $|s\bar{s}\rangle$ state, whereas such a
strange-quark content is hidden in the sea of the target proton.
Therefore, in contrast to the cases of $\rho^0$ and $\omega$,
little contribution  is expected from meson and baryon exchanges
to the $\phi$ meson photoproduction.
In Ref. \cite{titov}, one can
find more discussion on how to evaluate the strangeness component
in the proton contributing to the reaction process.

\begin{figure}[b]
\bigskip
\bigskip
\centering \epsfig{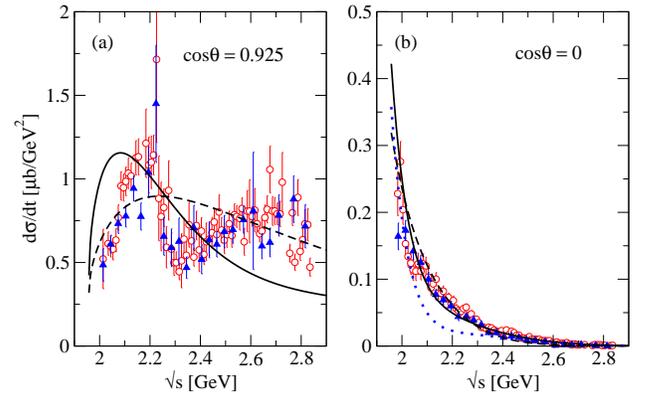}
\caption{Energy dependence of the differential cross sections for
$\gamma p\to \phi  p$ at the production angles
$\theta=22.33^{\circ}$ (a) and $\theta=90^{\circ}$ (b) in the c.
m. frame. The prediction of the present model given by the solid
line in (a) shows the role of the $\sigma$  exchange in the
observed peak at $\sqrt s\simeq 2.2$ GeV at the forward angle in
comparison to the black dashed line without the $\sigma$ exchange.
In (b), we show the dependence of the cross section on the phase
of the Regge pole at the angle $\theta=90^{\circ}$. The blue
dotted line results from the $\sigma$, $\pi$, and $f_2$ with all
canonical phases chosen. The solid and dashed lines are with and
without the $\sigma$ exchange, respectively, while  the complex
phase is taken for the $\pi$ exchange. Data are from
Ref.~\cite{dey2}.} \label{phi-3}
\end{figure}

In this respect, the results of the recent experiments by the
LEPS~\cite{mibe} and CLAS \cite{dey1} Collaborations are
interesting because a bump structure is observed in  the
differential cross section $d\sigma/dt$ around $E_\gamma\approx 2$
GeV (see Fig. \ref{phi-3}), which cannot be expected from the
monotonic behavior of the simple Pomeron-exchange model. Moreover,
as the author pointed out in Ref. ~\cite{dey2}, such a bump
structure gradually disappears as the scattering angle increases
from the forward to the mid-angle $\theta\approx 90^\circ$.  Of
course, one might immediately suspect the nondiffractive
subprocesses by the $\pi$ and $\eta$ exchanges in this region, but
their contributions are not enough to play the role.
In previous works on this issue, there were a few theoretical
attempts to account for the appearance of the bump structure by
including a nucleon resonance \cite{kiswandhi}, or the
$K^+\Lambda$(1520) coupled channel \cite{ozaki,ryu}   on the basis
of $\pi$, $\eta$, and the Pomeron exchanges.

In this work, motivated by the issue still open yet, we reexamine
the $\gamma p\to\phi p$ process with a focus on finding other
possibilities to explain the bump structure near threshold.
Based on the well-established result in the high energy realm
where the Pomeron exchange provides the diffraction in the
$t$-channel and the exchange of the tensor meson $f_2$ gives the
long-range contribution up to $\sqrt{s}\approx 10$ GeV, we here
investigate the role of the scalar $\sigma$ exchange in addition
to the $\pi$ exchange for the description of the reaction process
near the threshold. In existing model calculations, the former
exchange is usually excluded mainly because of the large
uncertainty in determining its coupling strength. Nevertheless, we
recall that the role of the $\sigma$ exchange is crucial to agree
with the peak of the total cross section observed in the $\gamma
p\to\rho^0\, p$ process \cite{ysoh2}. Furthermore, from the
well-known aspect that the natural parity exchange would dominate
the process of vector meson photoproduction, it is desirable to
consider the $\sigma$ exchange in the presence of the $\pi$
exchange, in particular, in the low energy region.

With this in mind, we discuss the possibility of the bump
structure driven by the $\sigma$ exchange in the $\gamma p\to\phi
p$ process. Differential and total cross sections are analyzed for
this purpose. Meanwhile, it is known that as the reaction energy
increases, the cross section for hadron interactions shows the
scaling as a manifestation of the quark structure of the hadron
\cite{brodsky}. In our previous study on the $\gamma p\to\pi^0\,p$
process, we discussed such an energy independence of the cross
section at the large transverse momentum transfer or alternatively
at the mid-angle $\theta=90^\circ$ \cite{kong}. Likewise, we may
well expect the scaling in the scaled differential cross section,
$s^7d\sigma/dt$, for the $\gamma p\to\phi p$ process at the
mid-angle in connection with the recent CLAS data~\cite{dey1}. The
scaled differential cross section is composed of three parts: the
resonance region where the cross section is governed by  hadronic
degrees of freedom, the scaling region in which quark and gluon
degrees of freedom are mainly involved, and the transition region
lying between them. Therefore, it will be interesting to see at
what energies the quark degrees of freedom start to show up in the
scaled differential cross section.
\\

The Reggeized amplitude for the $\gamma p\to\phi  p$ process
consists of the Pomeron ($\mathbb{P}$), $f_2$, $\sigma$, and $\pi$
exchanges which are given by
\begin{eqnarray}\label{neutral}
&&{\cal M}(\gamma p\to \phi p)={\cal M}_\mathbb{P}+{\cal
M}_{f_2}+{\cal M}_\sigma+{\cal M}_\pi\,,
\end{eqnarray}
where
\begin{widetext}
\begin{eqnarray}\label{charge0}
&&{\cal M}_\mathbb{P}=12i{e\,\beta_q\beta_{q'}\over f_\phi}
{m_\phi^2\over m_\phi^2-t}\left({2\mu_0^2\over
2\mu_0^2+m_\phi^2-t}\right)e^{-i{\pi\over
2}[\alpha_\mathbb{P}(t)-1]}\left({s\over
4s_0}\right)^{\alpha_\mathbb{P}(t)-1}F_1(t)\bar{u}(p')
(/\kern-6pt{k}\eta^*\cdot\epsilon-\rlap{/}\epsilon\eta^*\cdot
k)u(p)\,, \\
&&{\cal M}_{f_2}=\Gamma_{\gamma
f_2\phi}^{\beta\rho}(k,q)\Pi_{\beta\rho;\lambda\sigma}(Q)
\bar{u}(p')\Gamma_{f_2NN}^{\lambda\sigma}(p',p)u(p){\cal R}^{f_2}(s,t)\,,\\
&&{\cal M}_\sigma= \frac{g_{\gamma \sigma\phi}}{ m_0}g_{\sigma
NN}(k \cdot q\,\eta^*\cdot\epsilon-\epsilon\cdot q\,\eta^*\cdot
k)\, \bar{u}(p')u(p){\cal R}^{\sigma}(s,t)\,,\\
&&{\cal M}_\pi=i\,\frac{g_{\gamma\pi\phi}}{m_0}g_{\pi NN}
\varepsilon^{\mu\nu\alpha\beta}\epsilon_\mu\eta_\nu^* k_\alpha
q_\beta\,\bar{u}(p')\gamma_5u(p){\cal R}^{\pi}(s,t)\,,
\end{eqnarray}
\end{widetext}
with the Regge propagator,
\begin{eqnarray}
&&{\cal R}^{\varphi}(s,t)={\pi\alpha'_J\times{\rm
phase}\over\Gamma[\alpha_J(t)+1-J]\sin[\pi\alpha_J(t)]}
\left({s\over s_0}\right)^{\alpha_J(t)-J},\ \ \
\end{eqnarray}
written collectively for the $\varphi$ meson of spin-$J$ and
$s_0=1$ GeV$^2$. The phase of the $\varphi$ is, in general, taken
to be of the canonical form, ${1\over 2} [(-1)^J +
e^{-i\pi\alpha_J(t)}]$, unless it is exchange-degenerate. Here,
$\epsilon(k)$ and $\eta^*(q)$ are the photon and vector meson
polarizations with the momenta $k$ and $q$, respectively. Here,
$u(p)$ and $u(p')$ are the spinors for the initial and final
protons with the momenta $p$ and $p'$, respectively. Here,
$Q^\mu=(q-k)^\mu$ is the $t$-channel momentum-transfer.

The Pomeron exchange is expressed in terms of the quark loop
coupling in the $\gamma \mathbb{P}\phi$ vertex and $\mathbb{P}NN$
vertex with the nucleon isoscalar form factor given by
\cite{donnachie,ysoh}
\begin{eqnarray}
F_1(t)={4M^2-2.8t\over(4M^2-t)(1-t/0.7)^2}\,.
\end{eqnarray}
Since the Pomeron trajectory,
\begin{eqnarray}
\alpha_\mathbb{P}(t)=0.25\,t+1.08,
\end{eqnarray}
as well as the physical quantities such as the decay constant
$f_\phi=-13.4$, quark couplings $\beta_u=\beta_d=2.07$ GeV$^{-1}$,
$\beta_s=1.60 $ GeV$^{-1}$, and $\mu_0^2$=1.1 GeV$^2$ for the
quark loop in the $\gamma \mathbb{P}\phi$ are fixed to fit to data
at $E_\gamma\geq$ 10 GeV, we adopt these values without
modification.

The tensor meson $f_2$ exchange is expressed in terms of the
radiative coupling vertex given by~\cite{singer}
\begin{eqnarray}
&&\Gamma_{\gamma f_2\phi}^{\beta\rho}(k,q)=4\frac{g_{\gamma f_2
\phi}^{}}{m_0}\nonumber\\
&&\times(\eta\cdot\epsilon\, k^\beta q^\rho+k\cdot
q\,\eta^\beta\epsilon^\rho -\eta\cdot k\,\epsilon^\beta
q^\rho-\epsilon\cdot q\,\eta^\beta k^\rho),
\end{eqnarray}
and the tensor meson-baryon vertex,
\begin{eqnarray}%
\Gamma_{f_2NN}^{\lambda\sigma}(p',p)&=&\bar{u}(p')\Bigg[{2g^{(1)}_{f_2NN}
\over M}(P^\lambda \gamma^\sigma+P^\sigma \gamma^\lambda)
\nonumber\\
&+&{4g^{(2)}_{f_2NN}\over M^2}P^\lambda P^\sigma \Bigg]u(p)
\end{eqnarray}
with the spin-2 projection,
\begin{eqnarray}
\Pi^{\beta\rho;\lambda\sigma}(Q)&=&{1\over2}\left(
\bar{g}^{\beta\lambda}\bar{g}^{\rho\sigma}
+\bar{g}^{\beta\sigma}\bar{g}^{\lambda\rho}\right)
-{1\over3}\bar{g}^{\beta\rho}\bar{g}^{\lambda\sigma}.
\end{eqnarray}
Here $\bar{g}^{\beta\rho}=-g^{\beta\rho}+{Q^\beta Q^\rho\over
m_{f_2}^2}$, $P^\mu={1\over2}(p'+p)^\mu$, and $M$ is the nucleon
mass, and $m_0=1$ GeV. The coupling constant $g_{\gamma f_2
\phi}=0.0173$ is determined from the partial decay width $\Gamma_{
f_2\to\phi\gamma}=1.3$ keV~\cite{ishida} and the tensor
meson-nucleon coupling constants $g^{(1)}_{f_2NN}=6.45$ and
$g^{(2)}_{f_2NN}=0$ are taken from Ref.~\cite{bgyu}.

\begin{table}[t]
\caption{Listed are the physical constants and Regge trajectories
with phase factors for $\gamma p\to \phi p$. Here, $\varphi$
stands for $\sigma$, $\pi$, and $f_2$ of masses $m_\sigma=500$,
$m_\pi=134.9766$, and $m_{f_2}=1275.1$ MeV. For the coupling
constants associated with the $f_2 NN$ coupling, we use
$g^{(1)}_{f_2NN}=6.45$ and $g^{(2)}_{f_2NN}=0$.}
    \begin{tabular}{c|c|c|c|c}\hline
        Meson & Trajectory($\alpha_\varphi$) & Phase factor & $g_{\gamma \varphi \phi}$ & $g_{\varphi NN}$  \\
        \hline\hline
        $\sigma$ & $0.7(t-m_{\sigma}^2)$ & $(1+e^{-i\pi \alpha_{\sigma}})/2$ & $-0.085$ & 14.6 \\%
        $\pi$ &  $0.7(t-m_{\pi}^2)$ & $e^{-i\pi \alpha_{\pi}}$ & 0.065 & 13.4  \\%
        $f_2$ &  $0.9(t-m_{f_2}^2)+2$ & $(1+e^{-i\pi \alpha_{f_2}})/2$ & 0.0173 & 6.45; 0.0 \\%
        \hline
    \end{tabular}\label{tb1}
\end{table}

For the determination of $\gamma\sigma\phi$ coupling, it is
helpful to consider the partial decay width
$\Gamma_{\phi\to\pi\pi\gamma}$. We assume that the partial width
$\Gamma_{\phi \to \pi^0\pi^0\gamma}\approx 0.48$ keV in the
Particle Data Group (PDG) is mediated by the $\sigma$ meson and
obtain $g_{\gamma\sigma\phi}\approx 0.031$. On the other hand, we
note that Black {\it et al. }~\cite{black} predicted the partial
width $\Gamma_{\phi\to\sigma\gamma}=33$ keV based on the vector
meson dominance incorporated with the chiral effective Lagrangian.
This yields  $g_{\gamma\sigma\phi}\approx 0.146$ which is somewhat
larger than the naive evaluation from the $\pi\pi\gamma$ decay
width above. In the present calculation, we take the  value
$g_{\gamma\sigma\phi}=0.085$ which lies in the middle of the two
extremes.
The value of the $\sigma NN$ coupling constant in the literature
is very scattered and  found to be in a wide range of $5\sim
17.9$. We take $g_{\sigma NN}=14.6$ predicted by the QCD sum
rule~\cite{Erkol:2005jz,erkol} which is within the range of values
from the Nijmegen soft-core NN potential model~\cite{nagels}.

For the $\pi$ exchange, we take the coupling constants $g_{\pi
NN}$=13.4 and  $g_{\gamma\pi\phi}=0.065$ from the width
$\Gamma_{\phi\to\pi\gamma}=5.42$ keV reported in the PDG. In this
work, we choose the phase of $\pi$ exchange for a better agreement
with data. We summarize the coupling constants and Regge
trajectories with the phase factors in Table~\ref{tb1}. In the
calculation, the $\eta$  exchange as well as the scalar mesons
$f_0$ and $a_0$, axial meson $a_1$, and tensor meson $a_2$ are
neglected for simplicity because they appear in a minor role.

\begin{figure}[t]
\centering
\epsfig{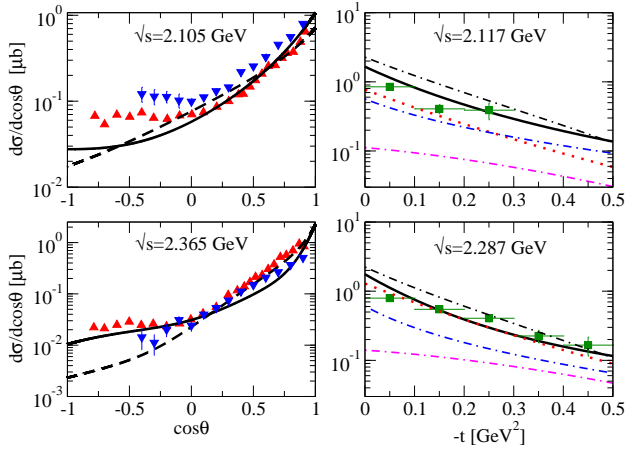} \caption{Differential
cross sections for $\gamma p\to \phi  p$  in the low energy
region. Left panels: The black solid and dashed lines  are the
cross sections with and without $\sigma$ exchange.  Right panels:
The contribution of each meson exchange is shown. The blue
dash-dotted line is from $\pi$ exchange, black dash-dotted line
from $\sigma$ exchange, magenta dash-dotted line from $f_2$, and
red dotted line from the Pomeron exchange. Data at
$\sqrt{s}=2.105$ and 2.365 GeV (red up triangle) are taken from
the charged mode \cite{dey1} and  at $\sqrt{s}=2.13$ and 2.38 GeV
(blue down triangle) from the neutral mode \cite{sera14} in the
CLAS Collaborations (left). The data in the right panel is from
the LEPS data~\cite{mibe}. } \label{phi-1}
\end{figure}

In Fig. \ref{phi-1}, we present the differential cross sections
for the $\gamma p\to\phi p$ process resulting from the coupling
constants and phases in Table \ref{tb1}. The role of the $\sigma$
exchange is illustrated in the CLAS data, and the contribution
from each meson exchange is shown in the LEPS data.

\begin{figure}[t]
\centering \epsfig{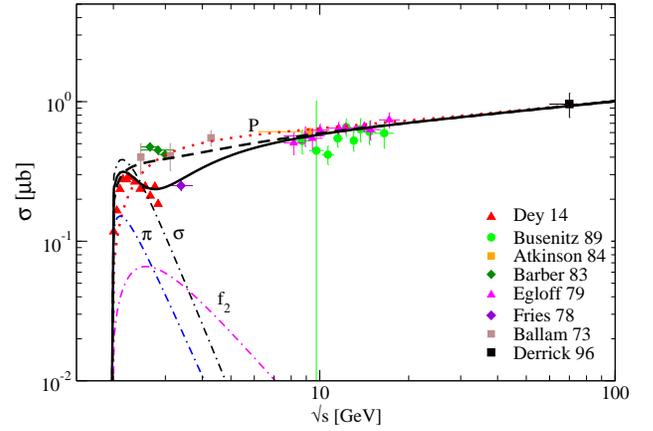}
\caption{Total cross section for $\gamma p\to \phi  p$ from
threshold to $\sqrt{s}=100$ GeV. Data are taken from
Refs.~\cite{dey1,derrick,busenitz,atkinson,Barber:1981fj,egloff,fries,Ballam:1972eq},
where the data points named as Dey 14 are obtained by integrating
over the differential cross sections given in Ref.~\cite{dey1}.
Our model favors the CLAS data~\cite{dey1} and data of Ref.
\cite{fries} as well. Notations for the curves are the same as in
Fig.~\ref{phi-1}.} \label{phi-2}
\end{figure}

Figure \ref{phi-2} shows the energy dependence of the cross
section from threshold up to the realm of the Pomeron exchange.
The data points near the threshold are obtained by integrating out
the data on the differential cross sections of Ref. \cite{dey1}.
As shown in the figure, there exists an inconsistency of the old
measurements~\cite{Barber:1981fj, Ballam:1972eq} with the recent
CLAS experiment~\cite{dey1}. Our result with the $\sigma$ exchange
as indicated by the black solid line agrees with the CLAS data,
whereas the result without it favors the old data as depicted by
the black dashed line. Thus, the cross section in the presence of
the $\sigma$ exchange  shows the feature quite contrasting to most
existing models. Indeed, in our model, the $\sigma$ exchange is
dominant near the threshold and, thus, plays the nontrivial role
to make the small peak around $\sqrt s\simeq 2.2$ GeV.

The role of the $\sigma$  exchange in the peaking behavior of the
cross section can be seen in other observables as well. Figure
\ref{phi-3} shows the dependence of the differential cross section
on the invariant energy $\sqrt{s}$ at two different angles of
$\phi$ production in the c. m. frame. Without the $\sigma$
exchange, the differential cross section in Fig. \ref{phi-3} (a)
describes nothing but the behavior passing through the average
value of data in the given energy range as shown by the dashed
line. In Fig.~\ref{phi-3}(b), we illustrate the dependence of the
differential cross section on the phases of the Regge poles
$\sigma$, $\pi$, and  $f_2$ at $\theta=90^{\circ}$. It should be
noted that the cross section with the canonical phases,
$(1+e^{-i\pi \alpha})/2$ for the $\sigma$ and $f_2$, and the
complex phase, $e^{-i\pi \alpha_{\pi}}$ for the $\pi$ exchange,
agrees with the experimental data, whereas the result with the
canonical phases for all the mesons shows considerable
disagreement.

Finally, let us discuss the possibility of observing the scaling
in the present process at the mid-angle. Brodsky {\it et
al.}~\cite{brodsky} predicted that the photoproduction cross
section obeys the power-law scaling, i.e.,
\begin{eqnarray}
    s^{n-2}{d\sigma\over dt}\sim F(t_0/s)\,,
\end{eqnarray}
for fixed $t_0$ based on the quark-counting rule. Here, $n=9$ is
the number of constituents (gauge boson plus the quarks)
participating in the $\gamma p\to\phi  p$ process. The measured
cross section is thus expected to exhibit such a scaling behavior
as $s^7{d\sigma/dt}\sim$ constant at the fixed angle around
$\theta=90^{\circ}$ (or fixed $t_0$) as energy increases.

\begin{figure}[]
\centering \epsfig{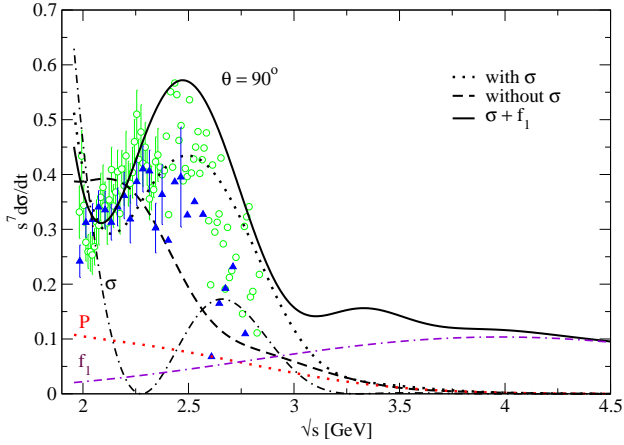}
\caption{Scaled differential cross sections $s^7{d\sigma\over
dt}$(10$^7$ GeV$^{12}$nb) for $\gamma p\to \phi  p$. The dotted
and dashed lines result from the calculation with and without the
$\sigma$ exchange, respectively, to exhibit its role crucial to
form the bump at $\sqrt{s}\simeq 2.5$ with the rapid drop
following before scaling begins. The solid and dotted lines are
from with and without $f_1$ in addition to the $\sigma$ exchange
to show that its role is substantial to manifest the scaling from
$\sqrt{s}\simeq 3$ to about 5 GeV. Here, $g_{\gamma
f_1\phi}=0.18$, the nucleon axial charge $m_A=1.08$ GeV, and
$\alpha_{f_1}(0)=0.9$.
The contributions of the relevant meson exchanges are denoted in
the figure legends. Data points at $\theta=90^\circ$ are obtained
from Ref.~\cite{dey2}.} \label{phi-4}
\end{figure}

Shown in Fig.~\ref{phi-4} is the scaled differential cross section
for $\gamma p\to\phi p$.  The data showed  the bump structure
around $\sqrt s \simeq 2.5$ GeV with a rapid drop following.
Two important points should be indicated in advance: the formation
of the bump by the $\sigma$ exchange before $\sqrt{s}\simeq3$ GeV
and the manifestation of the nonzero scaling by the $f_1$ exchange
above 3 GeV.
From the dotted and dashed lines for the scaled cross sections
with and without the $\sigma$ exchange, the role of the $\sigma$
exchange is crucial to drive such a bump structure with the nodes
around $\sqrt{s}\simeq 2.3$ and $3.3$ GeV which are given by the
vanishing of the canonical phase, i.e.,
$1+e^{-i\pi\alpha_\sigma(t)}=0$, and hence, $\alpha_\sigma(t)=-1$,
$-3$, and so on. Nevertheless, the scaled cross section from the
exchanges of  $\sigma+\pi+f_2+\mathbb{P}~$ approaches a vanishing
limit, and the energy independence of the cross section is not
clear as shown by the dotted line.

In order for the scaled cross section to manifest itself as being
a nonzero constant over the high energy region, interactions from
the quark-gluon dynamics are expected to contribute.

\begin{figure}[]
\centering \epsfig{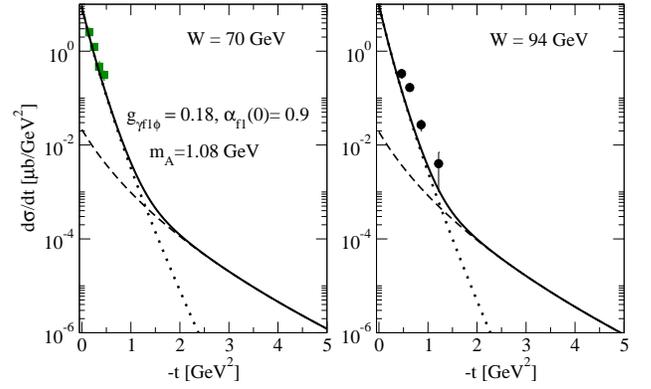} \caption{Role
of $f_1$ in the differential cross sections $d\sigma\over dt$ at
high energies. Solid and dotted lines are the calculation with and
without $f_1$ exchange, respectively. The dashed line shows the
contribution of the $f_1$ exchange. The role of $f_1$ is apparent
in the large $-t$ with $\alpha_{f_1}(0)=0.9$ with to respect Ref.
\cite{kochelev}. Data are from Refs.~\cite{derrick,breitweg}.}
\label{phi-5}
\end{figure}

In Ref. \cite{kochelev}, a new trajectory
$\alpha_{f_1}(t)=0.028\,t+(0.9\pm0.2)$ is suggested for the axial
vector meson $f_1(1285)$ of $1^{++}$ by relating  the properties
of $f_1$ with the two-gluon exchange via the axial anomaly of the
QCD vacuum. By considering the role of the $f_1$  peculiar to the
large $-t$ and energy, we calculate the contribution of the $f_1$
exchange with its role expected in a larger $-t$, i.e., a wider
range of the angle, as the energy increases. Over the region
$\sqrt{s}\simeq 3$ GeV, we obtain the scaling apparent to sustain
the energy independence up to 5 GeV due to the $f_1$ exchange
which is given by \cite{kochelev}
\begin{eqnarray}
&&{\cal M}_{f_1}=i{g_{\gamma f_1\phi}\over
m_0^2}m_\phi^2\epsilon_{\mu\nu\alpha\beta}k^\mu\eta^\nu\epsilon^\alpha
\left(-g^{\beta\lambda}+Q^\beta
Q^\lambda/m^2_{f_1}\right)\nonumber\\&&\hspace{0.5cm}\times
\left({1\over1-t/m^2_{A}}\right)^2 g_{f_1
NN}\overline{u}(p')\gamma_\lambda\gamma_5 u(p){\cal R}^{f_1}(s,t).
\end{eqnarray}
We use the canonical phase $(-1+e^{-i\pi\alpha_{f_1}(t)})/2$ and
the trajectory $\alpha_{f_1}(t)=0.028\,t+0.9$ which is within the
range of the intercept given in Ref. \cite{kochelev}. The cutoff
mass $m_A=1.08$ GeV is chosen for the nucleon axial form factor
\cite{liesenfeld} with  $g_{f_1 NN}=2.5$, and $g_{\gamma
f_1\phi}=0.18$ taken from the decay width
$\Gamma_{f_1\to\phi\gamma}= 0.019$ MeV reported in the PDG.
In practice, the physical quantities are applied to the
differential cross section in parallel with the scaled cross
section to cross-check the validity of those quantities for both
observables. The size of the cross section $s^7d\sigma/dt\approx
0.1$ [10$^7$ GeV$^{12}$nb] thus determined is consistent with the
differential cross sections as shown in Fig. \ref{phi-5}. It
should also be remarked that the contribution of the $f_1$
exchange is insignificant to other observables and could not alter
much the results we have shown above.

The limitation of the Regge trajectory on such a large angle as
discussed in Ref.~\cite{kong} is extended by the inclusion of the
$f_1$ exchange, and our results show some evidence of the special
role of the $f_1$ axial meson as advertised in
Ref.~\cite{kochelev}. Such a scaling obtained by the $f_1$ meson
exchange in this photoproduction of $\phi$ is quite in contrast
with the scaling-violating oscillatory behavior seen in the
proton-proton elastic scattering at a fixed angle~\cite{pire}. In
this sense, it is anticipated from future experiments to see
whether the scaling persists in photoproduction of the $\phi$
vector meson. We hope that there should be a measurement in the
region above 2.8 GeV in future experiments.

It is interesting to compare the present result with that from the
$s^{12}$-like scaling for the $\gamma p\to\phi p$ process in Ref.
\cite{dey3} where the number of gluons in the hadrons and photon
are counted more to give $n= 14$. The difference of the counting
numbers between the present work and Ref. \cite{dey3} leads to the
different energy region expected to scale; i.e., the expected
energy region the scaling appearing in Ref. \cite{dey3} is below
$\sqrt{s}\simeq 3$ GeV, whereas our model predicts the scaling to
start above 3 GeV.

In summary, we investigated the $\gamma p\to \phi p$ reaction
process with our interest in the possible role of the $\sigma$
exchange as the natural parity in the low energy region. Total,
differential, and the scaled cross sections are reproduced by the
$\sigma+\pi+f_2$ Regge poles on the basis of the background
contribution from the Pomeron exchange up to $\sqrt{s}=100$ GeV.
The role of the $\sigma$ exchange in addition to the $\pi$, $f_2$,
and the Pomeron exchanges is illustrated to account for the small
peak near the threshold in the total cross section and the bump
structure apparent in the  differential as well as the scaled
differential cross section. In this respect, the $\sigma$ exchange
is an important ingredient to understand the production mechanism
through the successful description of the observables we have
demonstrated in the present work.

With its role in the large $-t$ and energy by the new trajectory
arising from the axial-charge distribution of the QCD vacuum, the
exchange of the axial vector meson $f_1$ is exploited to clarify
the scaling above $\sqrt{s}\simeq3$ GeV in the scaled cross
section. In that region, where quarks and gluons are expected to
be involved, the result is positive to our expectation, i.e., the
QCD effect through the exchange of the $f_1$ trajectory
specialized to the QCD vacuum via the axial anomaly.

Viewed from the possibility of different powers of  $s^{n-2}$ as
well as the special role of the $f_1$ related to the QCD vacuum
via the axial anomaly, these findings in the scaling, in
particular, would deserve focus on the high-energy photon-beam
experiment to explore the quark-gluon dynamics in future
experiments, such as the LEPS2 at SPring-8 and CLAS12 planned at
JLab.


\acknowledgments
The work of B.-G. Yu was supported by Basic Science Research
Program through the National Research Foundation of Korea(NRF)
funded by the Ministry of Education, Science and
Technology(NRF-2013R1A1A2010504). The work of H. Kim was supported
by Basic Science Research Program through the National Research
Foundation of Korea(NRF) funded by the Ministry of Education(Grant
No. 2015R1D1A1A01059529).

\end{document}